\documentclass{ifacconf}

\usepackage{graphicx}      
\usepackage{natbib}        
\usepackage{amsmath}
\usepackage{amssymb}
\usepackage{graphicx}
\usepackage{caption}
\usepackage{subcaption}

\begin{document}
\begin{frontmatter}

\title{Graph Attention Inference of Network Topology in Multi-Agent Systems\thanksref{footnoteinfo}} 

\thanks[footnoteinfo]{© 2024 the authors. This work has been accepted to IFAC for publication under a Creative Commons Licence CC-BY-NC-ND”. This work was supported by the Strengthening Teamwork for Robust Operations in Novel Groups (STRONG) CRA at the Army Research Lab under the contract W911NF20-2-0089. }

\author[First]{Akshay Kolli} 
\author[First]{Reza Azadeh}
\author[Third]{Kshitij Jerath}

\address[First]{Richard Miner School of Computer and Information Sciences, University of Massachusetts Lowell, MA, USA (akshay\_kolli@student.uml.edu, reza@cs.uml.edu).}
\address[Third]{Department of Mechanical Engineering, University of Massachusetts Lowell, MA, USA 
    (kshitij\_jerath@uml.edu)}

\begin{abstract}                
Accurately identifying the underlying graph structures of multi-agent systems remains a difficult challenge. Our work introduces a novel machine learning-based solution that leverages the attention mechanism to predict future states of multi-agent systems by learning node representations. The graph structure is then inferred from the strength of the attention values. This approach is applied to both linear consensus dynamics and the non-linear dynamics of Kuramoto oscillators, resulting in implicit learning of the graph by learning good agent representations. Our results demonstrate that the presented data-driven graph attention machine learning model can identify the network topology in multi-agent systems, \textit{even} when the underlying dynamic model is not known, as evidenced by the F1 scores achieved in the link prediction.
\end{abstract}


\end{frontmatter}

\section{Introduction}
Networks are pivotal in modeling and understanding a wide range of natural and engineered complex systems (\citet{lewis2011network}, \citet{newman2018networks}) such as technological networks (\citet{cai2016}),  social networks (\citet{borgatti2009network}), biological systems (\citet{baronchelli2013networks}, \citet{sporns2018graph}). Within the context of complex multi-agent systems (MAS), modeling is enabled significantly by formulating their dynamics via a network representation. These networks define the inter-agent interactions as well as system characteristics such as the convergence properties in consensus dynamics or formation control, (\citet{olfati2007consensus}). To control, influence or understand such systems we require a deeper knowledge of their underlying network topology. Unfortunately, the network structure is often not available (or is only partially known) and must be inferred by observing the behavior of the multi-agent system, which is a challenging problem. 

Understanding the topology of systems where the system dynamics are unknown is useful across a myriad of systems biological networks and social networks. Identification of the emergence of team structures in coordinated tasks, evaluation of paths and connections in neuron, and graph analysis of gene regulatory networks all provide great insights without requiring the dynamics of the system to be known.

Several previous works have sought to solve the challenge of network topology inference, often also referred to as graph estimation or graph learning. Both \citet{ortega2018} and \citet{mateos2019connecting} provide an excellent survey of these efforts. These works have largely been focused on identifying the graph structure based on incoming datastreams from nodes without fully leveraging the underlying agent or node dynamics. On the other hand, the modeling and controls research community has focused a large part of its attention on control and coordination problems in networked systems where the underlying graph structure is either known or some appropriate assumptions can be made about it (\citet{wen2020}). This leaves a knowledge gap which, if addressed, can benefit both research communities: predicting the dynamics of individual agents or nodes can assist with the network topology inference challenges, and learning the underling interaction graph of dynamic agents can assist with control of networked multi-agent systems. 

Towards this end, some recent works have attempted to use graph signal processing and graph attention network approaches to address this knowledge gap (\citet{shuman2013emerging}, \citet{batreddy2023robust}). For example, \citet{zhu2020network} use a graph signal processing approach to recover the adjacency matrix using snapshots of consensus dynamics. Their work estimates the eigenvectors of the graph Laplacian using the sample covariance of node data and demonstrates that the spectral properties of the recovered graph matched those of the original network structure. To our knowledge, this work is limited to linear consensus problems. Similarly, from a graph attention network perspective, \citet{sebastian2023learning} present a machine learning model to extract information from node trajectories in a multi-agent system. This approach shows promise when applied to consensus dynamics and flocking scenarios. However, despite its innovative approach, this method requires knowledge of the true adjacency matrix for at least one example of the problem class, to be able to accurately predict the adjacency matrix. In our opinion, this presents a relatively strong assumption, and the approach is potentially unsuitable in settings where prior knowledge of comparable graphs is unavailable.

In contrast, the approach we present here is applicable to even those scenarios where no prior information of the network topology is available. Specifically, we propose a data-driven graph attention mechanism approach (\citet{velivckovic2017graph}) that learns the \textit{unknown graph structure} (unsupervised) while using full state observations from the \textit{unknown dynamics} of agents or nodes for a prediction task (supervised). We demonstrate this work not only in the linear consensus dynamics setting, but also for nonlinear synchronization scenarios such as Kuramoto oscillators (\citet{rodrigues2016kuramoto}). Our approach is able to concurrently learn to predict the future states of MAS exhibiting collective behavior, \textit{as well as} learn the underlying topology of the system within the same framework, as we demonstrate below.
\section{Problem Formulation}

We begin by considering a dynamic model of a networked system operating on graph $G = (V, E)$, where $V$ denotes the set of nodes (or agents) in the system, and $E$ denotes the edges of the network. 
The number of agents is defined as $|V| = N$. Let $A$ be the adjacency matrix of graph $G$, i.e. $a_{ij} \in \{0,1\}$ where $a_{ij} = 1$ indicates the presence of an edge between nodes $i$ and $j$, and $a_{ij} = 0$ denotes the absence of one.

We assume that the networked system has undirected edges, so that the adjacency matrix is symmetric, i.e. $a_{ij} = a_{ji}$ for all $(i,j)$.
The state of the $i^{th}$ agent at time $t$ is given by $x_i(t) \in \mathbb{R}^1$, and the collective state of the networked system is given by $\textbf{x}(t) = [x_1(t), x_2(t), ... , x_N(t) ]^\top$. The consensus dynamics are given by:
\begin{equation}
    \dot{\textbf{x}}(t) = -\mathcal{L}\:\textbf{x}(t)
    \label{eqn:consensus}
\end{equation}
where $\mathcal{L}$ represents the graph Laplacian. We also demonstrate our work using Kuramoto oscillators, for which the dynamics are given by:
\begin{equation}
    \dfrac{d\phi_i}{dt} = \omega_i + \dfrac{K}{N}\sum_{j = 1}^N \text{sin}\:(\phi_j - \phi_i) 
    \label{eqn:kuramoto}
\end{equation}

where $\phi_i(t)$ represents the phase of the $i^{th}$ agent (i = 1, 2, ... N) at time step $t$, $\omega_i$ denotes the innate frequency of the $i^{th}$ oscillator, $K$ represents the coupling constant. The collective state is $\textbf{x}(t) = [\phi_1(t), \phi_2(t), ..., \phi_N(t)]^\top$. The set of Kuramoto oscillators exhibits collective synchronization behavior as result of the coupling between them (\citet{strogatz2000kuramoto}).

The goal of this work is to develop a graph attention-based mechanism that can learn the network topology of a networked multi-agent system with unknown dynamics by learning to predict the state of the system (\citet{guo2019graph}). We build a machine learning model to obtain neural network parameters $\Theta^*$ such that:
\begin{equation}
    \Theta^* = \underset{\Theta}{\text{argmin }} 
    \mathbb{E}[ \text{loss} (\textbf{x}(t), \hat{\textbf{x}}(t)) ]
    \label{eqn:ml-model}
\end{equation}
where $\textbf{x}(t)$ is the state of the system at time step $t$ $\text{loss}$ is the mean absolute loss function, and $\hat{\mathbf{x}}(t) = f_\Theta(\{\mathbf{x}\}_{t-\tau:t})$ is the prediction of the state made by the learned neural network model at time step $t$.
The learned model $f_{\Theta}$ includes an interpretable attention layer or matrix $\hat{A}$ that we demonstrate can map on to the adjacency matrix $A$ as it represents the degree to which agents interact with one another. Once the model parameters have been learned, we can use the attention matrix $\hat{A}$ as an approximate representation of the adjacency matrix $A$ and thus can help inform us of the underlying topology that governs the dynamics of the multi-agent system.

\section{Learning the Network Topology}
\label{Sec:Estimating-the-network-topology}

The attention mechanism (\citet{bahdanau2014neural}) is a powerful mechanism that allows a model to dynamically pay attention to specific parts of its input (\citet{xu2015show}). 
Attention has found its use in Graph Neural Networks (\citet{guo2019graph}) providing a way for nodes to attend to each of their neighbors and producing node representations that perform well in downstream tasks. One downstream task that is especially useful for networked dynamic systems is the prediction of future state trajectories of agents in such systems. This task can find applications in various fields such as trajectory tracking (\cite{yang2024energyguideddatasamplingtraffic}), identifying agent influence (\cite{ jerath2024ZoneOfInfluence}), model predictive control, and reachability analysis, to name a few. We now discuss the architecture of the attention-based neural network model.

\begin{figure}[h]
  \centering
  \begin{subfigure}[b]{0.40\textwidth}
    \includegraphics[width=\textwidth]{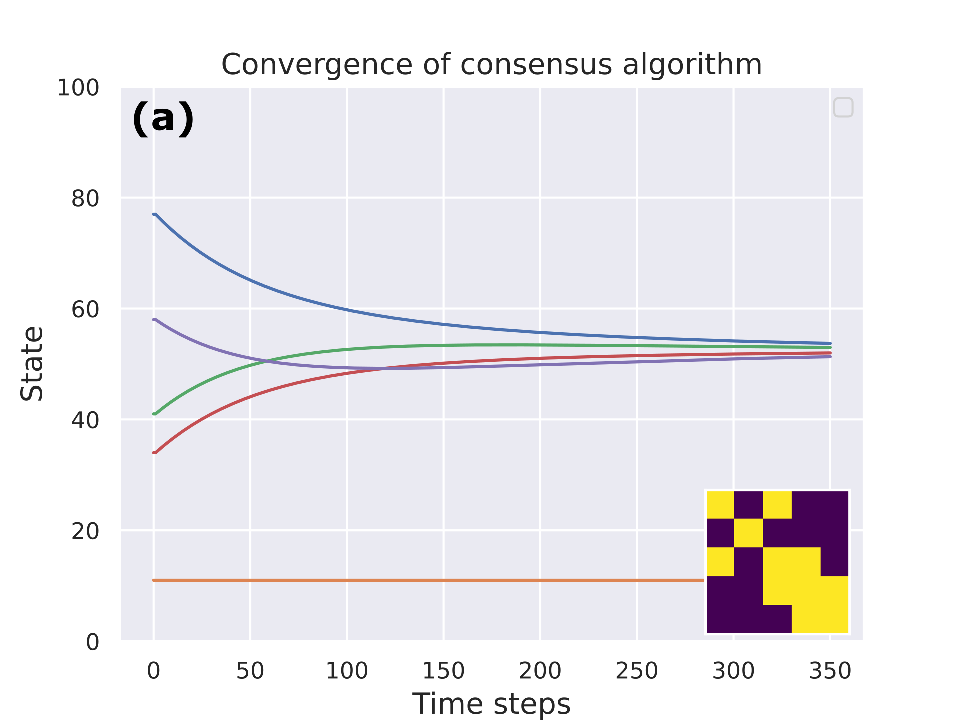}
    \label{fig:consensus_simulation}
  \end{subfigure}
  \hfill 
  \begin{subfigure}[b]{0.40\textwidth}
    \includegraphics[width=\textwidth]{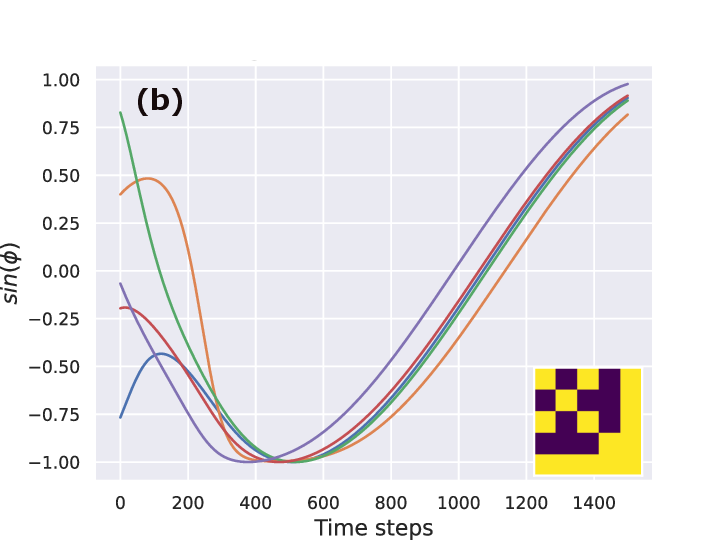}
    \label{fig:kuramoto_simulation}
  \end{subfigure}
  \centering
  \caption{Examples of the simulation dynamics. (a) Consensus dynamics simulation (b) Kuramoto Oscillator simulation. The y-axis on both the plots represents the state information for each agent. The x-axis represents the time steps taken in the simulations. The plots on the bottom right represent the graph that the simulations operate on, with yellow signifying a connection in the adjacency matrix, and purple representing a zero.}
  \label{fig:simulations}
\end{figure}

\subsection{Model Architecture}
\label{SubSec:3-Model-Architecture}
Our graph attention-based predictive model learns vector representations for each agent in the multi-agent system. In the process of learning to predict the states of the model, the model also learns the relationship between each of the agents and encodes it into their representations. Our novel contribution is the formulation and placement of the attention mechanism, such that attention scores are representative of the adjacency matrix in the multi-agent system, while simultaneously contributing to the predictive power of the model.

The attention mechanism in this model acts like a selective focus, enabling the model to dynamically prioritize and weigh the influence of different agents on each other. It operates by generating specific vectors—termed `key' and `query' vectors—from the initial representations of agents. These vectors help determine the extent of influence or attention one agent should pay to another during the learning process. This is achieved by comparing all pairs of key and query vectors across agents, effectively assessing their mutual relevance.

After establishing the relevance between agents, the model computes a set of attention scores. These scores are akin to the entries of an adjacency matrix in graph theory, depicting the connections between nodes (agents). Higher scores indicate stronger relationships or influences. These scores then modulate the interaction dynamics in the network by focusing the model’s learning on more influential connections, hence enhancing its predictive accuracy.

This attention mechanism is inspired by techniques originally developed for machine translation, where the model needs to decide which words in a source sentence are most relevant when predicting a word in the translation. Here, instead of words, the model evaluates which agents in a system are most relevant to predict future states effectively. This dynamic, context-based focusing capability is what enables the model to perform robustly in complex, variable multi-agent environments.

\begin{figure}
    \centering
    \includegraphics[width=8cm]{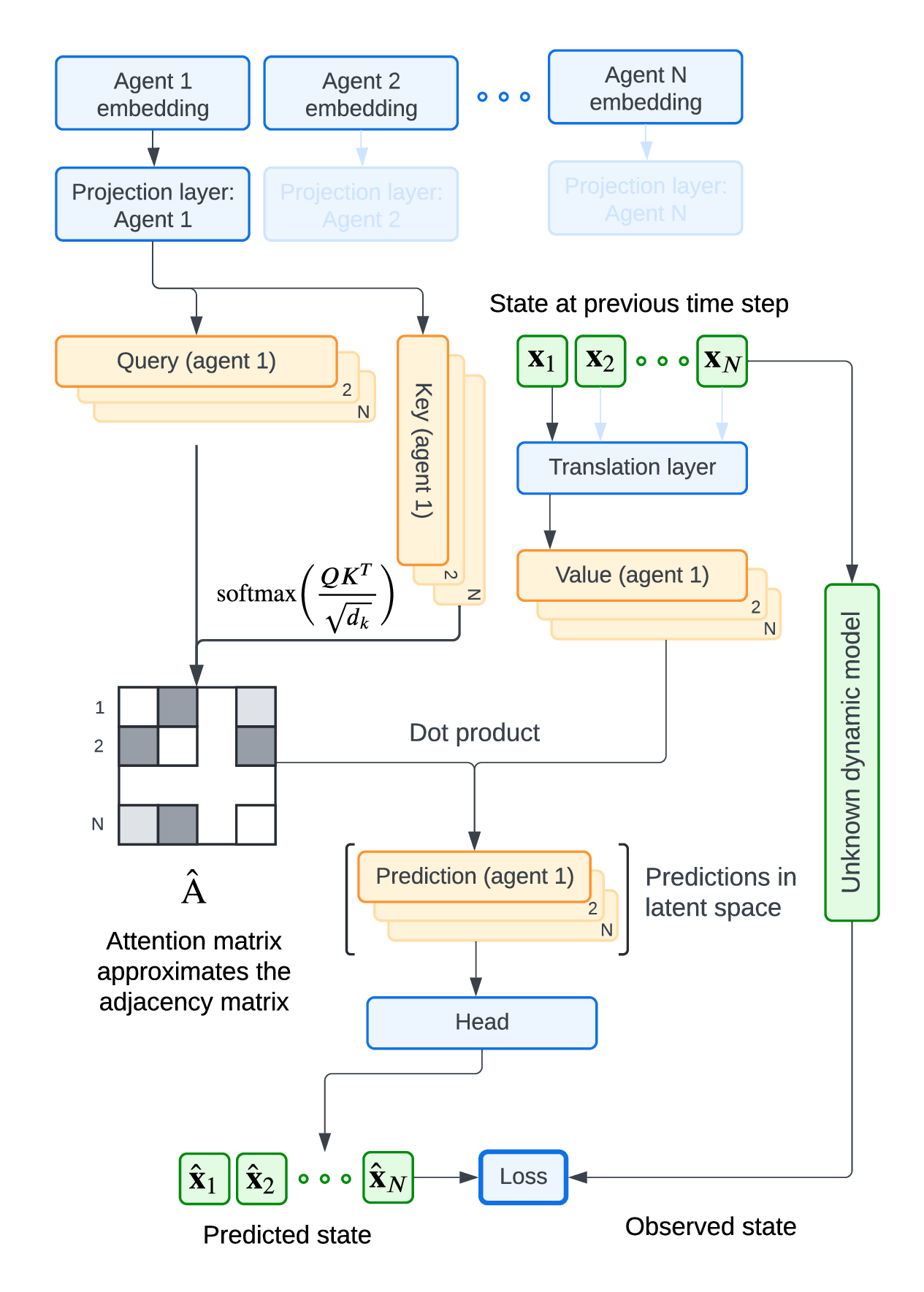}
    \caption{The model consists of 4 key components (shown in blue) that get learned: The agent embeddings, the translation layer, the attention projection layer and the head. }
    \label{fig:model_architecture}
\end{figure}

As illustrated in Figure \ref{fig:model_architecture}, the model consists of several neural layers, each communicating and passing information using vectors in the model's latent space. The model's latent space is a $d$-dimensional vector space. We set $d=64$ in all of our experiments. This setup ensures that the internal representations maintain a consistent level of complexity and detail. The model consists of four main components that are learned (indicated by blue boxes in Figure \ref{fig:model_architecture}), which are listed below. 

\vspace{-0.4em}
\noindent \textbf{Agent Embeddings:} These embeddings are a collection of vectors, each vector is a representation of an agent in the multi-agent system. The vector embeddings converge to being a good representation of each agent during training such that a high prediction accuracy of trajectories is achieved.

\vspace{-0.4em}
\textbf{Attention projection layer:} This neural network layer produces the key and query vectors from the agent embeddings. The key and query vectors are subsequently used to compute the attention between each of the agents.

\vspace{-0.4em}
\textbf{Translation Layer:} This neural network layer translates the state information of each agent into a vector in the model's latent space.

\vspace{-0.4em}
\textbf{Head:} This neural network layer converts a vector from the model's latent space into a prediction in the state space of the MAS.

Our method incorporates a technique first used in machine translation (\citet{vaswani2017attention}). The first step is to generate a key and a query vector from each of the agent embeddings using the projection layer. A value vector is generated from the state $\textbf{x}_i(t)$ for each agent using the translation layer. Scaled dot product attention is computed using the keys and the queries, giving us the attention matrix $\hat{A}$. The prediction is then obtained by passing the dot product of the attention matrix $\mathbb{R}^{N\times  N}$ matrix and the values $\mathbb{R}^{N \times d}$ through the head layer.

\subsection{Model Training}
\label{SubSec:3-Model-Training}

The model is trained using data generated from numerical simulations of the dynamic models for the multi-agent systems, i.e. the consensus dynamics and the Kuramoto oscillators. In these simulations, the underlying network topology for the multi-agent system is a randomly generated Erd\H{o}s-R\'enyi graph with edge probability $p = 0.5$. 

The agent states in consensus dynamics are initialized randomly within predetermined bounds, while those of the Kuramoto oscillators are randomly initialized between $(-\pi, \pi)$. Our experiments suggested that a larger initialization bound is helpful in preventing vanishing gradients during back-propagation. The simulations are run for up to 1000 timesteps to ensure that the dataset is dominated by the transient dynamics only, while avoiding steady state conditions which would limit the inference capabilities for this or any similar inference approach.

More specifically, the training dataset consists of the state $\textbf{x}(t)$ at one timestep as the input 
, and the state $\textbf{x}(t+1)$
at the next timestep as the target. Our choice of using state information at a single timestep as the basic constituent of the training data is driven by the need to ensure that the neural network learns the network topology. 
If longer sequences of state information were to be used as inputs, the neural network would learn to extrapolate the trajectory rather than infer the network topology.

During training, the model uses the current system state $\textbf{x}(t)$, processes it through the graph attention-based neural network, and obtains the predicted state $\hat{\textbf{x}}(t+1)$. This prediction is compared against the true states $\textbf{x}(t+1)$ known from the numerical simulations that constitute the training dataset. The resulting mean absolute error loss is then used to optimize the neural network model weights, and an ADAM optimizer is used to perform the back propagation.

Through the training process, the model refines its agent embedding vectors to better capture the characteristics of the agents. As the training progresses through the epochs, the attention between the agent embeddings $\hat{A}$ converges towards the true adjacency matrix $A$. In the next section, we discuss the efficacy of our approach in inferring the network topology as it learns to become proficient at predicting the future states of a dynamic system.

\section{Results}

To examine the performance of our approach, we first apply a threshold function to convert the real-valued attention matrix $\hat{A}$ to a binary-valued approximate adjacency matrix. The threshold value was heuristically selected to be $-0.4$ (based on the observed attention scores). Values in the attention matrix that were lower than the threshold were set to zero, i.e., low attention scores were replaced by 0 to indicate the absence of an edge between the agents, while higher attention scores were set to 1 (to represent the presence of an edge). Next, we use the F1 score to compare the binary-valued approximate adjacency matrix against the true adjacency matrix used for the numerical simulation of the dynamic system. The F1 score is a well-known measure for binary classification problems, and is defined as the harmonic mean of precision and recall. Specifically, the F1 score is defined as:
\begin{equation}
    F1 = 2\: \dfrac{\mbox{precision}\: \times \: \mbox{recall}}{ \mbox{precision + recall}}
\end{equation}
where precision is defined as the ratio of the number of correctly predicted links to the total number of correctly and falsely predicted links. It measures the accuracy of positive predictions.
On the other hand, recall, also known as sensitivity, measures the ability of a model to find all the relevant cases within a dataset. It is defined as the ratio of the number of correctly predicted links to the actual number of links in the network.

\subsection{Network Topology Inference in Consensus Dynamics}

\begin{figure*}[t]
  \centering
  \includegraphics[width=\linewidth]{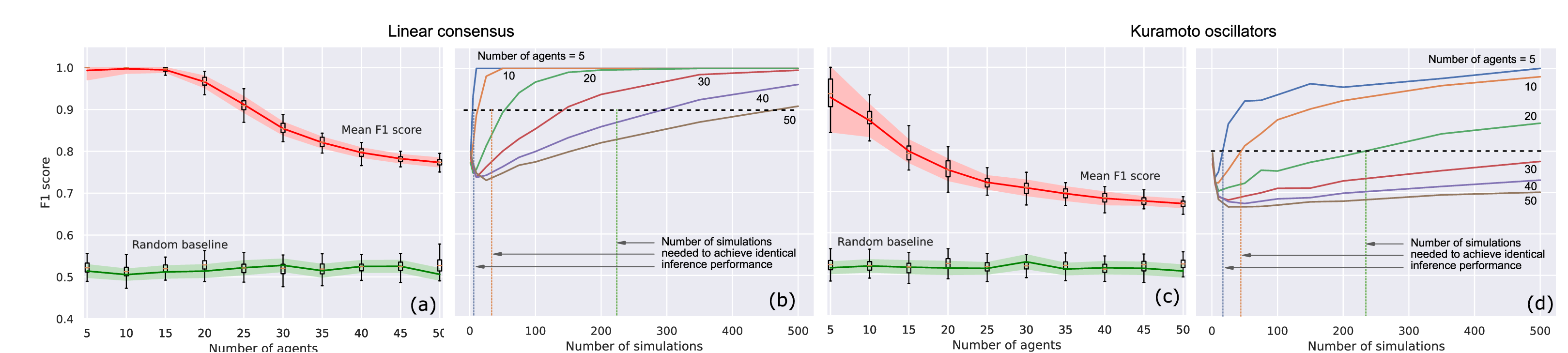}
    \vspace{-1em}
  \caption{(a), (c) Proposed method yields higher F1 scores (i.e., better performance) for graph inference with small number agents in both Consensus dynamics and Kuramoto oscillators. If data is limited, performance drops for systems with more agents. Performance calculated with 100 simulations worth of data. (b), (d) Using additional simulation data improves inference performance for larger systems as well.}
  \label{fig:results}
\end{figure*}

 Figure \ref{fig:results}a shows the F1 score performance of the model for systems with varying number of agents in the multi-agent system. The green line serves as a baseline, representing the mean F1 score achieved by adjacency matrices generated from random graphs in comparison to the target graph. The model shows a gradual decline in performance as the number of agents increases. Each training run had n=100 simulation runs. The decrease in performance could be because of the larger number of relationships to learn, causing it to require more than 100 simulation runs to accurately represent the adjacency matrix. Another limitation could be the use of the softmax functions, which obscures weaker connections as the number of agents increases.

 Figure \ref{fig:results}b shows the F1 score plotted against the number of simulations in the training set. The different lines represent multi-agent systems with increasing number of agents. To make equally accurate graph topology inference, systems with a small number of agents require considerably fewer number of simulations in the training data than systems with larger number of simulations. Though systems with a large number of agents start of with a significantly lower F1 score, the score steadily increases with increase in the number of simulations. The quality of predictions for a systems of any size are limited by the training resources available.

 \subsection{Network Topology Inference in Kuramoto Oscillators}

Figure \ref{fig:results}c shows the plots for the F1 score performance of the model in predicting the adjacency matrix for a system of Kuramoto oscillators. The accuracy for systems of all sizes are worse, when compared to their consensus dynamics counterparts. The inherent non-linear nature of Kuramoto oscillators makes it more difficult for the trained model to provide accurate predictions. Nevertheless, the performance does improve with increase in the number of simulations, as evidenced in Figure \ref{fig:results}d, making the training resources be the limiting factor for achieving better performance.

\subsection{Learning of attention throughout training}

\begin{figure*}[t]
    \centering
    \includegraphics[width=1.2\linewidth]{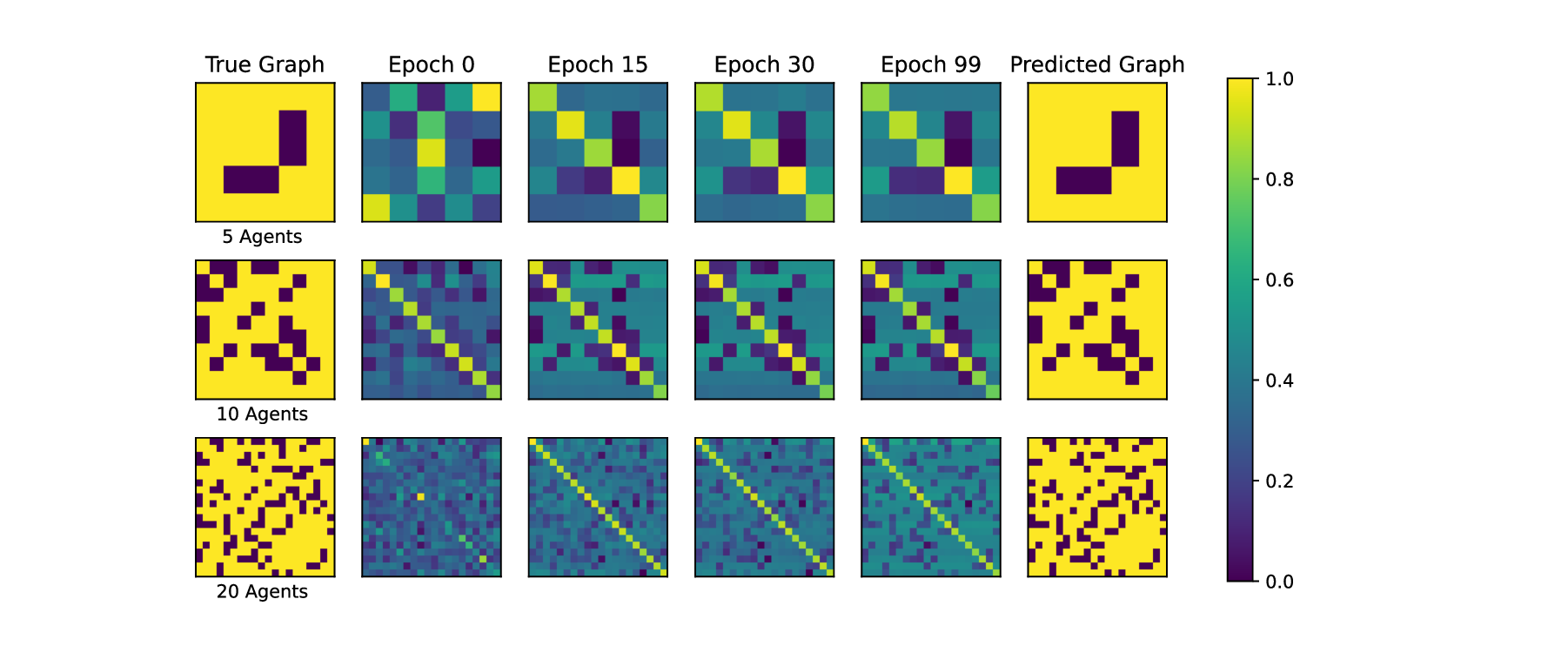}
    \vspace{-3em}
    \caption{Visualizing attention values through training stages, with the final column thresholding the attention value in the final epoch to give the predicted adjacency matrix. Reults demonstrate that most of the learning is done through the early epochs, with the later epochs adding to the final details.}
    \label{fig:learning_over_time}
\end{figure*}

Figure \ref{fig:learning_over_time} shows the attention that is learned during the different stages of training. Each agent pays the most attention to itself, which is intuitive, since each agent's future state is most likely to correspond with its current state. This is the easiest relationship to learn, hence the first that is learned. The remaining relationships are learned gradually over time, but remain weaker than the attention being paid to themselves.

\section{Concluding Remarks}

In this work, we tackled the problem of \emph{learning the graph that a multi-agent system operates on, from the state history of the system.} 

Multi-agent systems are often found to be non-linear in nature, and studying the model performance on Kuramoto oscillators can extend our understanding of synchronization phenomena, highlight the importance of network topology in collective dynamics, and offer insights into designing more efficient and robust distributed systems for various applications. We demonstrated that pure machine learning approaches can be used to construct flexible powerful models for graph prediction on linear and non-linear systems. The key contribution of this work is to present a graph learning or network topology inference approach for multi-agent systems where both the prior structure of graphs as well as system dynamics is unknown. They provide flexibility in that they can be used for a large array of problems. The underlying dynamics need not be known to be able to predict the graph. Possible future applications can be in measuring trust or reliance on teammates in autonomous multi-agent teams, dynamic adjustment of roles in teams optimizing team performance, coordination in decentralized clusters, and improving Human-Robot interaction.

Our findings align with recent studies that emphasize the importance of prioritizing agents based on their specific properties \citet{Findik2023} and incorporating social interactions into multi-agent systems \citet{Hossein2021}, reinforcing the potential of relational awareness and network topology in guiding cooperation strategies and shaping team behaviors.

\subsection{Future Work}

In our study, we delved into the application of our system to undirected graphs, demonstrating its potential for straightforward extension to directed graphs. The next phase of our research should aim at refining the system to accurately extract edge weights along with the topology, thereby enhancing its applicability to weighted directed graphs.

We anticipate extending our research to the dynamic identification and tracking of time-varying graphs. As the underlying graph evolves, so too should the learned representations, enabling the continuous detection and monitoring of these dynamic changes. This progression holds significant promise for the advancement of graph-based learning systems and their application across a myriad of complex, real-world scenarios.

\bibliography{ifacconf}             

\end{document}